\def\go{\raisebox{-.2ex}{$\raisebox{-.3ex}{$>$}\atop
\raisebox{.3ex}{$\sim$}$}}
\documentstyle[12pt,a4]{article}
\begin{document}
\begin{titlepage}
\begin{flushright}
	USITP-94-03\\
      PAR-LPTHE 94-06\\
      hep-ph/9403239\\
\end{flushright}
\vskip .8cm
\begin{center}
{\Large
Gamma ray lines from TeV dark matter}
\vskip .8cm
 {\bf L. Bergstr\"om}\\
\vskip .5cm
Department of Stockholm\\
Stockholm Univeristy\\
Box 6730, S-113 85 Stockholm\\
Sweden
\vskip .3cm
and\\
\vskip .3cm
 {\bf J. Kaplan}\\
\vskip .5cm
Laboratoire de Physique Th\'eorique et Hautes Energies\\
Universit\'e de Paris VI et VII\\
Unit\'e associ\'ee au CNRS (UA 280)\\
2 pl. Jussieu, F-752 51 Paris\\
France
\vskip 1.8cm
\end{center}
\begin{abstract}
We calculate, using unitarity, a lower bound on the branching ratio
$\chi\chi\to \gamma\gamma$ and $\chi\chi\to \gamma Z$,
 where $\chi$ is any halo dark matter
particle that has $W^+W^-$ as one of the major annihilation modes.
Examples of such particles are supersymmetric particles with a dominant
Higgsino component, or heavy triplet neutrinos. A substantial
branching ratio is found for the $\gamma\gamma$ and $\gamma Z$ modes.
We estimate the
strength of the monoenergetic $\gamma$ ray lines that result from such
annihilations in the Galactic or LMC halos.
 \end{abstract}
\end{titlepage}
\newcommand{\beq}{\begin{equation}}
\newcommand{\eeq}{\end{equation}}

The question of the extent and nature of the dark matter in the Universe
 continues to be one of the most pressing problems
of contemporary physics and astrophysics. Large scale structure and
peculiar velocity
observations seem to favour a value of the energy density much larger than
what baryons can contribute due to nucleosynthesis constraints (for a
recent
review, see \cite{darkmatter}). The recent candidate detections of
microlensing
events in the halo \cite{machos} are, although very interesting, not
conclusive as concerns the baryonic content of the dark halo of our galaxy
(see, e.g., \cite{turner}). Even if a large population of dark compact
objects
(such as brown dwarfs) were formed at an early stage of galactic evolution,
the overdensity of baryonic matter thus formed would accrete surrounding
cold
dark matter creating a particle dark matter halo density at least
comparable
in magnitude to the baryonic one.

One of the  favoured particle dark matter candidates is the lightest
supersymmetric particle $\chi$, assumed to be a neutralino, i.e. a mixture
of
the supersymmetric partners of the photon, the $Z^0$ and the two neutral
$CP$-even Higgs bosons present in the minimal extension of the
supersymmetric standard model (see, e.g. \cite{haberkane}). The
attractiveness
of this candidate stems from the fact that its generic couplings and mass
range
naturally gives a relic density close to the critical one. Besides, its
motivation from particle physics has recently become stronger due to
the apparent need for 100 GeV - 10 TeV scale supersymmetry to achieve
unification of the gauge couplings in view of recent LEP results
\cite{amaldi}.

When it comes to detecting cold dark matter particles, it seems that
the technology is not yet advanced enough for the ultrasensitive detectors
developed to register the nuclear recoil and/or ionization to put
interesting
bounds on supersymmetric dark matter. On the other hand, indirect
detection methods look quite promising. With large neutrino telescopes
like DUMAND \cite{dumand} and AMANDA \cite{amanda} now being deployed,
there is a fair chance to detect energetic neutrinos from the center
of the Sun or  Earth coming from annihilations of captured
neutralinos, if they constitute the dark matter halo \cite{stelzer}.
In fact, some bounds have already  been obtained from the much smaller
Kamiokande detector \cite{kamiokande}.

The other way to detect particle dark matter in the halo is through
indirect
detection of positrons, antiprotons and $\gamma$ rays generated through
the continuous annihilation of dark matter particles in the halo
 \cite{indirect}. Observations of $\gamma$ rays have the advantage of
giving a sensitive map of the galactic halo (since the annihilation rate
depends on the square of the local dark matter density). In particular,
if there is an enhancement of the density at the center of our galaxy
\cite{ipser,venya1,silkstebbins}
or the Large Magellanic Cloud \cite{paolo}, the $\gamma$ ray flux could
stand out well above background.

A particularly interesting annihilation process in the halo is
$\chi\chi\to\gamma\gamma$ or $\chi\chi\to Z^0\gamma$. Since these
are two-body final states and the annihilating massive particles move with
non-relativistic speed in the halo (typically $v/c\sim 10^{-3}$) the
produced $\gamma$s will be nearly monoenergetic, meaning a $\gamma$
ray line signature \cite{gammalines}. The calculations of
$\chi\chi\to\gamma\gamma$ for
$m_\chi<m_W$ have been made in quite some detail \cite{smallmass,venya}.
The
result
is that the signal is only marginally detectable with present-day space
detectors \cite{venya}. On the other hand, it has recently been pointed
out that a very massive $\chi$ in the TeV region could give a detectable
signal in Air Cherenkov Telescopes (ACT) on the ground \cite{urban}. These
detectors have a very large effective area (on the order of 20 000 $m^2$)
and
a good proton rejection and energy resolution ($\sim 10$ \%) can be
obtained
with modern techniques.

In \cite{urban} the $\gamma\gamma$ process was estimated for the case of a
nearly pure Bino, based on earlier
calculations. However, the rates turn out to be very small
\cite{urban,error}.
 There
remains to do the more difficult calculation for  a Higgsino, both for
the $\gamma\gamma$ and the $Z^0\gamma$ final states. The latter,
which to our knowledge is considered for the first time here, will in
fact turn out to be the most important one for Higgsinos (and for any
other dark matter candidate that has a strong coupling  to the $W^+W^-$
final state in its  annihilations). Although a full calculation
remains to be done for, say, the minimal supersymmetric extension of the
standard model, we will use unitarity to put a strict lower bound on the
$\gamma$ line signal as well as an estimate of the full rate.

We consider first the Higgsino annihilation $\chi^0\chi^0\to W^+W^-$,
mediated by
chargino exchange in the $t$ and $u$ channel  (there are also $Z$ and $H$
exchange contributions in general, but these vanish in the $v\to 0$ limit).
The calculation is identical for any Majorana particle that annihilates
to $W$ pairs through the exchange of a charged fermion.
We write the $\chi^0\chi^\pm W^\mp$ coupling (see \cite{GKT,haberkane} for
conventions; the index 0 here identifies the lighter of the two charginos)

\beq
\Gamma^\mu={ig\over 2}\gamma^\mu(f_0P_L+e_0P_R),
\eeq
where $P_{L,R}=(1\pm\gamma^5)/2$.

In the limit where the lightest neutralino is a pure heavy Higgsino,
the $\chi\chi\to~WW$ amplitude is dominated by the exchange of the
lightest chargino, nearly degenerate with the neutralino, and
\beq
|e_0|=|f_0|=1
\eeq
Projecting out the $S$ wave part of the initial amplitude by the
projector
\cite{KKS}
\beq
{\cal O}_{Ps}=-{m_\chi\over \sqrt{2}}\gamma^5(1-{p_\chi\over m_\chi}),
\eeq
we find the effective $\chi\chi W^+W^-$ vertex for polarization indices
$\mu$ and $\nu$ of the $W$ bosons with four-momenta $p_+$ and $p_-$:
\beq
V_{\chi\chi W^+W^-}=\biggl( {g^2(e_0^2+f_0^2)\over
2\sqrt{2}(m_{\chi^0}^2+m_{\chi^+}^2-m_W^2)}\biggr)
\epsilon[\mu\nu p_- p_+],\label{effective1}
\eeq
where $\epsilon_{\mu\nu\rho\sigma}$ is the completeley antisymmetric
constant
tensor, and we use the abbreviation $\epsilon[\mu\nu p_- p_+]\equiv
\epsilon_{\mu\nu\rho\sigma}p_+^\rho p_-^\sigma$.
This gives the annihilation rate
\beq
\sigma v (\chi\chi\to W^+W^-)={g^4\over 128 \pi m_\chi m_W}
{(\omega-1)^{3/2}(e_0^2+f_0^2)^2\over (1-\omega-\kappa)^2},
\eeq
where $\omega=(m_{\chi^0}/m_W)^2$, $\kappa=(m_{\chi^+}/m_W)^2$.
This result coincides with the $v\to 0$ limit of the expresion given in
\cite{GKT}.

We are now prepared for the one-loop calculation of $\chi\chi\to
\gamma\gamma$. We assume $m_{\chi^+}\go m_{\chi^0}>>m_W$. It has been
shown in a similar type of calculation \cite{lars-rad} that it is a very
good approximation, even when $m_{\chi^+}\sim m_{\chi^0}$, to take the
effective vertex (\ref{effective1}) as pointlike :
\beq
V_{\chi\chi WW}\sim C\epsilon[\mu\nu k (k-q_1-q_2)],\label{effective2}
\eeq
with
\beq
C={g^2(e_0^2+f_0^2)\over
2\sqrt{2}m_{\chi^+}^2}
\eeq
(see Fig.~1). For the  contribution of the imaginary part to the
branching
ratio which we will extract to obtain a lower bound on the cross section,
expression (\ref{effective2}) is even
exact because the denominator of the chargino propagator is constant.
One can easily convince oneself that the direct
$\chi\chi
 WW\gamma$ vertex (Fig.~1(b)) that by gauge invariance has to be present
due to the
 derivative coupling in the effective vertex (\ref{effective2}) does not
 contribute to our process. In addition, the ordinary  contact
 $WW\gamma\gamma$  term (Fig.~1(c)) contributes zero, since the intitial
low-velocity
 $\chi\chi$ state only projects out the part of the amplitude
antisymmetric in the $W$
 momenta, and the contact term is symmetric in its indices and thus
 vanishes upon contraction.

 There remains to calculate the triangle graphs with $W$s and in the
 customary linear $R_\xi$ gauge also
 Goldstone bosons circulating  in the loop (Fig.~1(a)). Considerable
simplification
is obtained by choosing a non-linear gauge condition as originally
suggested
by Fujikawa \cite{fujikawa}. This has been used in various applications
such as calculating Higgs decays into two  photons
\cite{gavela,lars-hulth}, the charge radius of the neutrino
\cite{monyonko} or radiative neutralino decay \cite{wyler}.  The idea is
to
replace the usual gauge fixing term (we set $\xi=1$ for simplicity)
\beq
{\cal L}_{g.f.}=-{1\over 2}\left(\partial_\mu
A^\mu\right)^2-{1\over 2}\left(\partial_\mu Z^\mu+m_Z G^0\right)^2 -
|\partial_\mu W^{+\mu}+im_WG^+|^2,
\eeq
where $G^{\pm,0}$ are the Goldstone bosons eaten by $W^\pm$ and $Z^0$,
by
\beq
{\cal L}_{g.f.}^{n.l.}=-{1\over 2}\left(\partial_\mu
A^\mu\right)^2-{1\over 2}\left(\partial_\mu Z^\mu+m_Z G^0\right)^2 -
|(\partial_\mu +igW_\mu^3)W^{+\mu}+im_WG^+|^2,\label{nl}
\eeq
where $W_\mu^3$ is the neutral component of the $SU(2)_L$ gauge triplet
($W_\mu^3=Z_\mu^0\cos\theta_w - A_\mu\sin\theta_w$).
The main advantage of this gauge is that the new contribution from
(\ref{nl}) to the
trilinear $\gamma W^\pm G^\mp$ part of the Lagrangian actually cancels
a corresponding piece
of the original trilinear sector of the Lagrangian giving a vanishing
total
$\gamma W^\pm G^\mp$ coupling. In addition, the new terms entering the
$\gamma W^\pm W^\mp$ and $Z^0 W^\pm W^\mp$ vertices, although making them
superficially look more complicated (and less symmetric in the momenta of
the three bosons) in reality make these vertices much simpler when at
least one of the bosons is on  mass shell.
The diagram in Fig.~1(a) (and
the
one obtained by crossing the photon lines which actually gives an
identical
contribution) were calculated in the `t Hooft-Feynman version ($\xi=1$) of
the nonlinear gauge (in fact, we also checked our results using the more
cumbersome linear gauge).

We find the amplitude for $\chi\chi\to \gamma(q_1)\gamma(q_2)$ with photon
polarization
four vectors $\epsilon^\mu(q_1)$ and $\epsilon^\nu(q_2)$\ to be
\beq
{\cal A}_{\gamma\gamma}^{\mu\nu}={-8ie^2C\over (2\pi
)^4}\int d^4k{m_\chi^2\epsilon[k\mu\nu(q_1+q_2)]+
k_\mu\epsilon[k\nu q_1q_2]-k_\nu\epsilon[k\mu q_1q_2]\over
(k^2-m_W^2)((k-q_1)^2-m_W^2)((k-q_1-q_2)^2-
m_W^2)},
\eeq
where the ordinary prescription $m_W^2\to~m_W^2-i\epsilon$ is
understood in the  denominator.

Using standard techniques, this integral can be transferred to a symmetric
integral in $D=4-2\epsilon$ dimensions after combining the factors in the
denominator using Feynman parametrization. The $D$-dimensional integrals
are
of two types, one convergent and one logarithmically divergent as $D\to
4$ (signalled by an $1/\epsilon$ divergence as $D\to 4$).
The (infinite) renormalization constant is real and does not affect the
imaginary
part, which can be calculated in terms of elementary functions since the
2-dimensional Feynman parameter integral becomes 1-dimensional after using
the $\delta$-function coming from the $i\epsilon$ term in the
denominator. (This corresponds to using the Cutkosky rules to extract the
imaginary part of the amplitude.)
Writing
\beq
{\cal A_{\gamma\gamma}}={Ce^2\over 4\pi}\epsilon[\mu\nu q_1q_2]
\left(M_{re}+iM_{im}\right),\label{matrix}
\eeq
we find
\beq
M_{im}=\beta^2\log\left({1+\beta\over 1-\beta}\right),
\eeq
where $\beta=\sqrt{1-(m_W/m_\chi)^2}$.

In the limit of large $\chi$ masses (above, say, 500 GeV) this becomes

\beq
M_{im}\sim \log\left({4m_\chi^2\over m_W^2}\right).
\eeq
We have checked this result using still another method of calculation that
relies on
the generalized optical theorem, writing (the integral is over the $WW$
phase space)
\beq
M_{im}={1\over 2}\int M_{\chi\chi\to WW}M^*_{\gamma\gamma\to WW},
\eeq
where this time we employ the chargino propagator  rather than
the pointlike approximation for
the $\chi\chi WW$ vertex.
This gives the same result and thus in addition proves the assertion that
for
the
calculation of the imaginary part, the approximation of a pointlike
effective
vertex is exact whenever $m_{\chi^\pm}>m_\chi^0$.

Once the imaginary (absorptive) part has been found, the real (dispersive)
part
can be calculated using dispersion relation techniques. However, for the
pointlike vertex a
subtraction
will be needed, corresponding to a constant term resulting from the
chargino
propagator cutoff in the diagrams that generated
the effective vertex. For very large $\chi$ masses the leading
logarithmic
terms will dominate and can be calculated:
\beq
M_{im}^{LL}=\log({s\over m_W^2}),
\eeq
\beq
M_{re}^{LL}={1\over 2 \pi}\log^2({s\over m_W^2}),\label{real}
\eeq
where $s=4m_\chi^2$.

In the following we will use the
imaginary part to get a strict lower bound for the rate estimates. To get
an idea of the full result we will use the leading log estimate
(\ref{real}) for the real part.

For the application we are con\-sidering, anni\-hilation of heavy
non-\-rela\-tivistic
dark matter particles in the galactic halo, the $\gamma$ ray line coming
from $\chi\chi\to\gamma\gamma$ cannot realistically be discriminated from
the corresponding line from $\chi\chi\to Z^0\gamma$. For example, the
$\gamma$ energy for $m_\chi = 1$ TeV is 1 TeV for the $\gamma\gamma$ mode
and $(4m_\chi^2-m_Z^2)/(4m_\chi)=998$\ GeV for the $Z^0\gamma$ mode,
i.e. within the $10^{-3}-10^{-2}$
spread of velocities of the annihilating particles (and of course well
within the
$5-10$ \% energy resolution that at most can be achieved with present-day
Air Cherenkov Telescopes). We should therefore add the $\gamma$ ray
luminosities from the two sources. In fact, as we now shall see, the line
strength from $Z^0\gamma$ is expected to be a few times larger that that
from the $\gamma\gamma$ final state.

In the gauge we have chosen and to leading logarithmic accuracy,
the difference between the $\gamma\gamma$ and
$Z^0\gamma$ calculations  is simply the replacement $e^2\to
e^2\cot\theta_w$ in Eq.(\ref{matrix}). In the rate, this amounts to an
enhancement by
the factor $\cot^2\theta_w\sim 3.4$. When calculating the intensity of the
line
signal, the fact that each $\gamma\gamma$ event gives two photons is
compensated by the division of the symmetry factor ($=2$) due to the
presence of two identical particles in the final state.

In Fig.~2 we show our lower bound for the average number of $\gamma$
line photons of energy
$\sim m_\chi$ per $W^+W^-$ annihilation:
\beq
F_\gamma\equiv{2\sigma v(\chi\chi\to \gamma\gamma)+
\sigma v(\chi\chi\to Z^0\gamma)\over
\sigma v(\chi\chi\to W^+W^-)}=\alpha_{e.m.}^2\left(1+\cot^2\theta_w\right)
\log^2\left({4m_\chi^2\over m_W^2}\right),\label{flux}
\eeq
as well as the value for this rate resulting from the adoption
of (\ref{real}) for the dispersive part of the amplitude.  When
converting this into an effective line flux per $\chi\chi$ annihilation
one has to correct by a  factor giving  the branching ratio of
$\chi\chi\to W^+W^-$ normalized to the total annihilation cross section.
This typically means a reduction of a factor of $2$ for a Higgsino (due
mainly to the $ZZ$ channel), but
no reduction for a species (e.g. a triplet neutrino \cite{salati}) that
does not couple
to $Z$ bosons at tree level. The remarkable feature of our result
(\ref{flux}) is its very large size compared to the case of, e.g. a pure
Bino \cite{urban,error}.
 As can be seen in Fig.~2,
 the effective branching ratio to line photons can easily be as large as
 $10^{-2}$. There will of course also be a continuous, diffuse flux of
lower
 energy $\gamma$s coming from the fragmentation of the $W^+W^-$ and $ZZ$
 final states (see \cite{urban}). This distribution however lacks
 conspicuous features and has its main contribution at very low energies
 where the background $\gamma$ flux is much more severe and, what is
 worse, its shape is largely unknown.  It may therefore prove very
 difficult to detect dark matter candidates through this "soft" photon
 flux.
 If detectors
 (e.g. ACTs) customized to achieve high energy resolution are built, a
 line signal would be an obvious feature to search for in the data. If
 such a line is found, there is certainly no known astrophysical
 background that could account for it, and the energy would simply
 correspond to the mass of the dark matter particle.

 Although we have focused on the Higgsino, one of the presently favoured
 dark matter candidates, our results are in fact of much greater
 generality for the following reasons. Any massive (TeV scale) neutral
cold
 dark matter fermion may be expected to have a substantial annihilation
 strength into $W^+W^-$ pairs (although exceptions can be found, such as
 pure Binos). The fermion can be of Majorana or Dirac nature. For
 Majoranas, our analysis can be directly applied. The difference with
 Dirac fermions is that the CP properties do not forbid the slow fermions
 in the halo to annihilate in the triplet $S$ (vector) state. However, due
to
 gauge invariance, a massive spin 1 state can not decay into two photons,
 so the only contribution comes from the singlet $S$ state which is the
 one we have already treated. Thus, we expect the logarithmic enhancement
 to be a quite general feature. The actual branching ratio line photons
 may, however,  be lower by a factor of a few due to the new triplet
 contributions to the total annihilation rate.

In the case of a heavy ($m_N > m_W$) weak triplet Majorana neutrino (which
was considered in \cite{salati} for $m_N < m_W$), our results  need no
modification, since $WW$ is the only important tree level annihilation
mode. That model is an interesting example of a viable particle dark
matter model where direct detection is essentially impossible (since the
neutrino does not have tree level $Z$ couplings), but where $\gamma$ ray
lines are predicted at a substantial level.

We now turn to some estimates of the rates for the gamma line processes.
Since
the annihilation rate depends on the square of  the neutralino number
density,
the rate and angular distribution of the  $\gamma$ line flux will depend
on the
model for the Galactic halo. In  particular, there are some arguments that
there
may be a core region of  the galaxy where the dark matter density is
substantially higher than  average \cite{ipser}. In the model by
Berezinsky et al \cite{venya1}  the  density profile falls as
$1/r^{1.8}$ all the way from 5 Mpc
down to the  innermost part of the galaxy, where a black hole is thought
to be
residing. This model gives an enormous enhancement (several orders of
magnitude)
of the $\gamma$ ray flux  from $\chi$ annihilations in
the central region of the galaxy. Assuming this halo density profile,
already existing measurements of high energy cosmic gamma ray fluxes
constrain very strongly the supersymmetric models \cite{venya2} (for a
recent criticism of this model
for the halo, see \cite{flores}).

On the more conventional side, it has  recently
been pointed out \cite{paolo}
that the Large Magellanic Cloud may have its own halo (albeit with only
partially
known parameters), which possibly could give an enhanced flux from that
direction. We will consider these  cases, noting as in \cite{urban} that
for a smooth halo the rates we find are too small to be detectable with
present technology. On the other hand, one of the the main advantages
 for turning the
attention to the $\gamma$ ray signatures is just that this is a promising
way of mapping the density distribution of the halo in the case it is not
completely homogeneous. For instance, there is a possibility that string
or texture generated
perturbations in the cold dark matter density could have gone non-linear
early and survived tidal disruptions to provide a very clumpy halo
\cite{silkstebbins}.

For masses up  to a few TeV, we can neglect galactic absorption of the
$\gamma$ rays. According to \cite{venyabook} the optical depth at the
10 kpc length scale is around $10^{-3}$. (Remember that for the high mass
range
we are considering, even for the $Z\gamma$ channel, $E_\gamma=m_\chi$ to
an
accuracy of within a percent.)

For  a localized source of dark matter annihilations at the center of the
galaxy, the gamma line flux is given by
\beq
F_\gamma={(\sigma v)_\gamma\over r_\odot^2}
\int_0^{r_0}\left({\rho(r)\over m_\chi}\right)^2r^2dr,\label{fgamma}
\eeq

where $r_\odot\sim 8.5$ kpc is the distance of the solar system from the
center of the galaxy, and $\rho(r)$ is the density profile within the core
radius $r_0$.
For  a generic dark matter particle that gives a substantial contribution
to $\Omega$ of the universe, $(\sigma v)_{tot}$ is of the order of
$10^{-26}$ cm$^3$s$^{-1}$. Since the annihilation rate
of neutralinos into $W^+W^-$  does
not vanish nor become small when $<v> \to 0$, $(\sigma v)_{tot}$ is
expected to be of the
same order of magnitude in the galactic halo. As an example, this
happens for a Higgsino of
the minimal supersymmetric extension of the Standard Model of a mass
around 1 TeV. Since we see from Fig.~2 that the number of line photons
per annihilation is roughly of the order of a percent, we can use the
value $(\sigma v)_{line}\sim 10^{-28}$ cm$^3$s$^{-1}$ in our estimates.

In fact, for the case of a pure Higgsino, we can calculate the absolute
value of the lower bound of
the line cross section. In that case $f\,=1$, $e_0=\pm 1$ in
Eq.\,~(1), and we find (assuming near Higgsino-chargino degeneracy, which
is generally an excellent approximation for a nearly pure, massive
Higgsino in
 minimal supersymmetric models)

\beq
2\sigma v(\chi\chi\to \gamma\gamma)+
\sigma v(\chi\chi\to Z^0\gamma)>
 {\pi\alpha_{e.m.}^4\over 8m_\chi^2\sin^6\theta_w}
\log^2\left({4m_\chi^2\over m_W^2}\right),
\eeq
which corresponds to $0.6\cdot 10^{-28}$ cm$^3$s$^{-1}$\ for $m_\chi
= 1$ TeV. Assuming, lacking a full calculation, a contribution from the
dispersive
part of the amplitudes of the same order of magnitude, we see that in this
case
the annihilation rate into line photons is indeed around  $10^{-28}$
cm$^3$s$^{-1}$

Using the Ipser-Sikivie model for the core of the halo \cite{ipser},
 a flux of
around  200 photons per year is predicted in an ACT of area 20 000 m$^2$
during a
typical observation "year" of $2\cdot 10^6$ $s$, with an estimated (but
poorly known) background about an order of magnitude lower (including
misidentified protons and electrons; see \cite{urban} for further
discussions). This rate for Higgsino-like particles is two orders
of magnitude larger than that expected for pure Binos \cite{urban,error}.
In the
model by Berezinsky et al \cite{venya1}, the flux can be still at least
two orders
of magnitude larger. However, if one assumes a  smooth halo distribution
$\rho(r)=\rho_\odot(r_\odot^2+a^2)/(r^2+a^2)$  with $a\sim r_\odot$, the
rate
falls below one event per year in our  generic ACT detector in the
direction of
the galactic center. In the LMC  model of Gondolo \cite{paolo} the rate
could be
enhanced by an order of  magnitude making it just about reaching the
detectability limit. It should  be noted, though, that the high threshold
(200-300 GeV) of present day  ACTs make these processes rate limited.
Going to
larger areas and lower  thresholds could greatly increase the discovery
potential
of this type of  detector.

To conclude, we have shown that the $\gamma\gamma$ and $Z\gamma$
annihilation rates  are quite high for any dark matter candidate that
couples with full strength to $W^\pm$. A typical case is provided by a pure
Higgsino, for which we have shown that the $\gamma$ line strength is very
much
larger than that of a pure Bino. It should be noticed, though, that our
calculations indicate this large $\gamma$ line rate to be generic for any
dark matter candidate that couples to $W^\pm$ with ordinary electroweak
strength. If the  galactic halo contains  regions with higher dark matter
density than the local (solar neighbourhood) value, the detection of high
energy
monoenergetic photons could at the same time determine the mass of the
dark
matter particle and map its galactic density distribution.

This research has been sponsored by the Swedish Natural Science Research
Council and a European Community Twinning grant. We wish to thank P.
Gondolo, F. Martin and H. Rubinstein for useful discussions. We are
especially grateful to M. Urban, whose idea to detect $\gamma$ lines from
WIMP
annihilation using  Atmospheric \v Cerenkov Telescopes
triggered this work.

\newpage
\vskip .2cm
\parindent 0pt
{\Large Figure Captions}
\vskip .3cm
\begin{itemize}
\item[1.] Some diagrams relevant to the annihilation $\chi\chi\to
\gamma\gamma$. For slow Majorana particle annihilation, only the diagram
in (a) contributes (the diagrams obtained by crosing the photon lines are
not shown in the figure).
\item[2.] The average number $F_\gamma$ (Eq.\,~(\ref{flux}))
 of $\gamma$ line photons from $\chi\chi\to
\gamma\gamma$ and $Z^0\gamma$ normalized to the $W^+W^-$ annihilation
rate, as
a function of the $\chi$ mass. The solid line is the lower bound obtained
by using only the imaginary part of the amplitude which is given by
unitarity. The dashed line is an estimate of the full rate obtained by
taking the leading logarithmic result of Eq.~(\ref{real}) for the real
(dispersive) part of the amplitude.
\end{itemize}

\begin{thebibliography}{99}
\bibitem{error} Note that there is an error in Eq.~(20) of \cite{urban},
entirely due to one of the present authors (JK).
The prefactor should read $8\times 10^{-31}$ cm$^3$s$^{-1}$ instead of
$4.2\times 10^{-29}$ cm$^3$s$^{-1}$.
\end{thebibliography}
\end{document}